\documentstyle[12pt,titlepage]{article}
\input epsf
\def\NPB{{\em Nucl. Phys.} B}
\def\PLB{{\em Phys. Lett.}  B}
\def\PRL{\em Phys. Rev. Lett.}
\def\PRD{{\em Phys. Rev.} D}

\def\be{\begin{equation}}
\def\ee{\end{equation}}
\def\bea{\begin{eqnarray}}
\def\eea{\end{eqnarray}}
\def\laq{\raise 0.4ex\hbox{$<$}\kern -0.8em\lower 0.62
ex\hbox{$\sim$}}
\def\gaq{\raise 0.4ex\hbox{$>$}\kern -0.7em\lower 0.62
ex\hbox{$\sim$}}

\begin{document}
\titlepage
\begin{flushright}{BGU-PH-96/08}\end{flushright}
\vspace{1in}
\begin{center}
\large{\bf SPECTRUM OF \\ COSMIC  GRAVITATIONAL WAVE BACKGROUND\\}

\vspace{2cm}
\normalsize
{ \bf RAM BRUSTEIN\\ }
\vspace{.5cm}
{\it Department of Physics,  Ben-Gurion University \\ Beer-Sheva, 84105, 
ISRAEL}
\vspace{2cm}


{\bf Abstract}
\end{center}

\noindent
Models of string cosmology predict a stochastic background of
gravitational waves with a spectrum that is strongly tilted towards high
frequencies.   I give simple approximate expressions for spectral densities
of the cosmic background which can be directly compared with
sensitivities  of gravitational wave detectors.

\vspace{1.8cm}
\noindent
\rule[.1in]{13.5cm}{.002in}\\

\noindent
 Contribution to the proceedings of the international conference
on gravitational waves: sources and detectors, Cascina (Pisa), 
Italy, March 19-23, 1996.

\newpage
\baselineskip=18pt

A class of  string cosmology models in which the evolution of the Universe
starts with a dilaton-driven inflationary phase, followed by a high
curvature (or ``string") phase  and then by standard radiation and matter dominated 
evolution was presented \cite{1}, and shown to predict a cosmic gravitational
wave {\small (GW)} background of characteristic type. Using general arguments, 
 numerical estimates of spectral parameters were improved \cite{bgv}. The
purpose of this talk is to provide simple approximate formulae for the spectrum
of the cosmic GW background which could be used for direct comparison with
measurements of present and planned experiments.  More details about the general
framework, the models and the computation of the spectrum  as well as
 previous relevant work can be found in  [3-10].

The  energy density today in {\small GW} at frequency $f$ in a bandwidth
equal to $f$,  $\rho_{G}(f)=
\frac{\hbox{ $d E_{G}$}}{\hbox{ $d^3 x\ d\ln(f)$}}$, 
produced during a dilaton-driven inflationary phase  was computed in 
[1] and is given, approximately, by
\begin{equation}
\rho_{G}(f)=\rho_{G}^{S}\left(y_S\right)
\left(\frac{f}{f_S(z_S)}\right)^3, \hspace{0.5in} f\le f_S.
\label{ogwdd}
\end{equation}
The two parameters of the model are {\small 1)} $y_S$, the ratio of
the string coupling parameter at the beginning and at the end of the high curvature
 phase, which
 is taken  to be $y_S\laq 1$,  and {\small 2)}
$z_S$, the ratio of the scale factor of the Universe  at 
 the end and at the beginning  of the string phase. The parameter $z_S\ge 1$ may
take very large values. The two parameters $y_S$, $z_S$
may be traded for the two parameters $\rho_{G}^{S}(y_S)$ and $f_S(z_S)$.

The  energy density in {\small GW} produced during the string phase cannot be
computed at present because of our inadequate understanding of high
curvature dynamics. However, we may boldly extrapolate the spectrum using 
 a single power to obtain the following spectrum  
$\rho_{G}(f)=\rho_{G}^{S} \left(\frac{f}{f_S}\right)^{2\beta}$ 
for $f_S\le f \le f_{1}$ where
$\beta=-\frac{\ln y_S}{ \ln z_S}>0$ and $f_{1}$ is the end-point frequency,
today, of the amplified spectrum. Note that  $\rho_{G}(f_1)$ is the maximal
energy density $\rho_{G}^{max}$.

We turn now to discuss  numerical estimates for the pairs 
$\left(f_{1},\rho_{G}^{max}\right)$ and $\left(f_{S},\rho_{G}^{S}\right)$. To
that end it is useful to consider the ``minimal spectrum", in which the the
dilaton-driven inflationary phase connects almost immediately to FRW
radiation dominated evolution. For the minimal spectrum $z_S=1$, $y_S=1$,
$f_{1}=f_S$ and  $\rho_{G}^{max}=\rho_{G}^{S}$.
The end-point frequency today  $f_{1}(t_0)$,  was red-shifted from its value
at the onset  of the radiation era $f_1(t_r)$ due to the expansion of the
Universe. Using entropy balance  we may evaluate the amount of
red-shift in terms of ratios of temperatures and effective numbers of degrees
of freedom. To evaluate $f_1(t_r)$ we  use energy balance 
at $t=t_r$  to relate $T(t_r)$ and the Hubble parameter at $H(t_r)$ and a
geometrical relation between the end-point frequency and $H(t_r)$, $f_1= 
H(t_r)/2\pi$. 
To determine  $\rho_{G}(f_{1})$ for the minimal spectrum we may use
the ``one-graviton" criterion, identifying the crossover from accelerated 
to decelerated  evolution
 when only one graviton per phase space cell is produced
$\rho_{G}(f_{1})= 16\pi^2 f_1^4$. The result is the following  range, 
depending on various parameters \cite{bgv}  
\begin{eqnarray}
f_{1}=f_S&=&0.6-2\times10^{10} Hz  \nonumber \\
\rho_{G}^{max}=\rho_G^S&=& .03-3 \times 10^{-6} \rho_c h_{50}^{-2}.
\label{ftodayfinal}
\end{eqnarray}
In the last equation the critical energy density $\rho_c$ and the 
Hubble parameter in units of $50\ km/sec/mpc$ were introduced.
Equations (\ref{ftodayfinal})  define  together the coordinates of the minimal
spectrum. As  will be shown elsewhere  using S-duality symmetry
arguments, the minimal spectrum is also a lower bound on the  amount of GW
energy density produced during the dilaton-driven phase.
For more details on the determination of 
$\left(f_{1},\rho_{G}^{max}\right)$ see [2].  

At the moment,  the most restrictive (indirect) experimental  bound on the 
spectral parameters comes  from  nucleosynthesis
(NS)  constraints \cite{ns1} $\rho_{G}\laq 0.1 \rho_R$, 
 $\rho_{G}(f_{1}) < 5\times 10^{-5} \rho_c h_{50}^{-2}$. As can be seen from
eqs.(\ref{ftodayfinal}), the nucleosynthesis bound is 
satisfied quite well. Previous, and very recent analysis of relevant direct 
and indirect  bounds on $\rho_G$  can be found in [12].

For non-minimal  spectra $f_S=f_{1}/z_S$ and  $\rho_{G}^{S}= \rho_{G}^{max}
y_S^{2}$, and the spectrum has a break at $f_S$ as in figure 1. The position
of the break $(f_S,\rho_G^S)$ covers a wedge in the $(f,\rho_G)$ plane.  
We  obtain estimates similar to   eq.(\ref{ftodayfinal}) on
$\rho_{G}^{S}$ and  $\rho_{G}^{max}$ and  summarize
the results   in  figure 1, describing  
the interesting wedge in
the  $(f,\rho_G)$ plane,  bounded in between the NS bound and the minimal
spectrum.\\

%\begin{figure}
%\rule{5cm}{0.2mm}\hfill\rule{5cm}{0.2mm}
%\vskip 2.0in
%\rule{5cm}{0.2mm}\hfill\rule{5cm}{0.2mm}
\centerline{\epsfxsize=3.2in\epsfbox{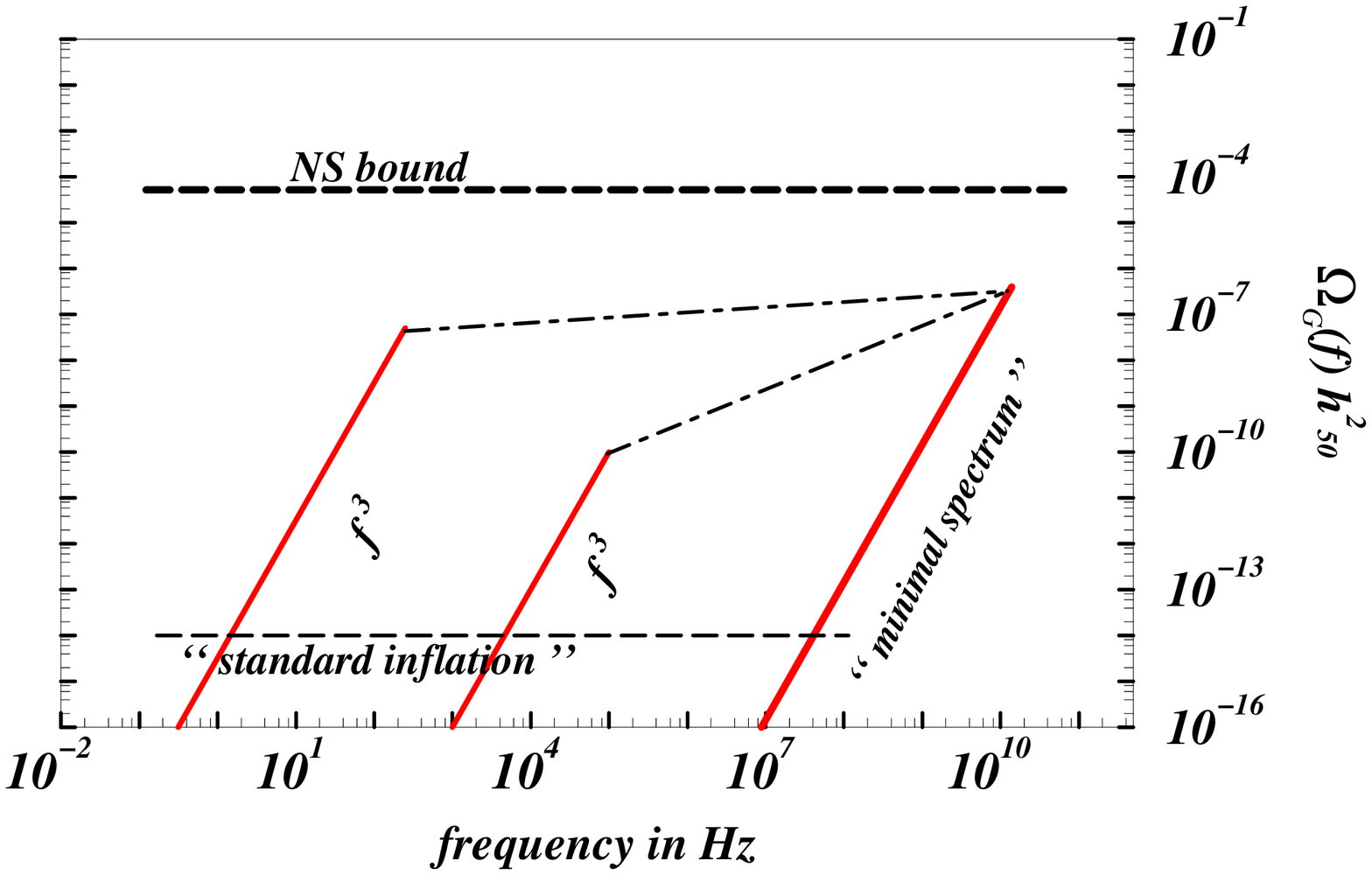}
\epsfxsize=3.2in\epsfbox{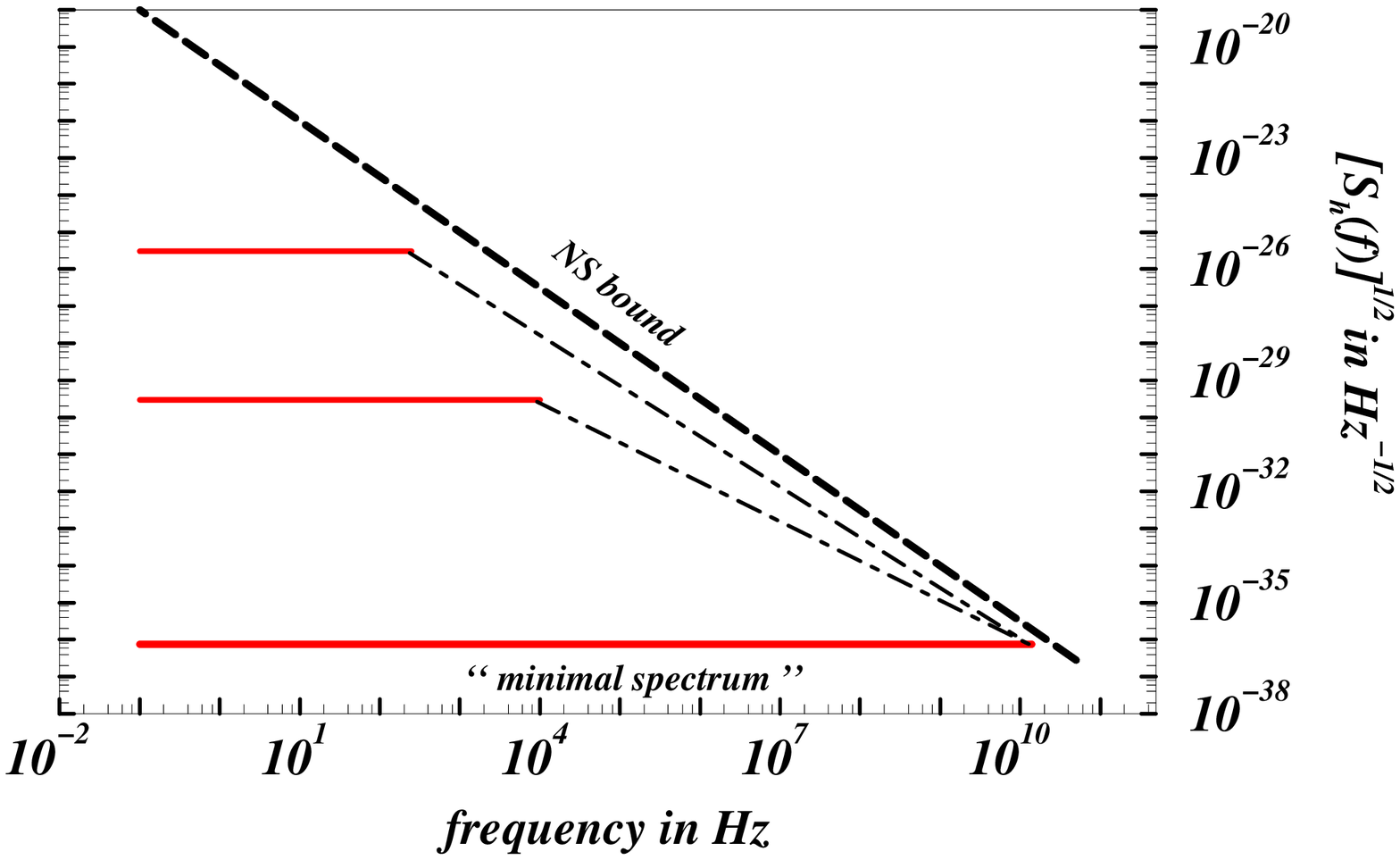}}

\hsize=12cm\hspace{0.01cm}
\vbox{{\noindent
\small\it \baselineskip=2pt
{\bf Figure 1.} Energy density $\Omega_G(f)=\rho_G(f)/\rho_c$
(left) and spectral amplitude (right) of gravitational waves (GW). 
The solid lines are possible spectra of  GW 
produced during the dilaton-driven phase. 
The minimal spectrum provides a lower bound
on energy density in  GW produced during the dilaton-driven phase
and the thick dashed line marks a nucleosynthesis upper 
bound. The dot-dashed lines are 
extrapolations of the   spectrum into the string
phase. The spectrum expected from slow-roll inflation is shown (left) for 
comparison.\hfil\vfill}}
\baselineskip=18pt
\hsize=\textwidth

Most {\small GW} experiments present their sensitivity in terms of the
spectral density $\sqrt{S_h(f)}$ of a metric perturbation $h$ rather than that of
$\rho_{G}(f)$. We will translate the previous results and present the corresponding
$S_h(f)$.  The two-sided  spectral density $S_h(f)$  is
given by 
$\langle\widetilde h (f_1) \widetilde h^*(-f_2)\rangle=
\frac{1}{2} \delta(f_1+f_2) S_h(f_1)$ 
where $\widetilde h ({x},f)$ is Fourier transform of each of the two 
physical, transverse-traceless metric perturbations and  $\langle\cdots\rangle$
denotes time or ensemble averages.  The average {\small GW} energy density for
isotropic and homogeneous background such as ours is given by
$
\frac{\hbox{ $d E_{G}$}}{\hbox{ $d^3 x$}}=\frac{1}{8 \pi G_N}
\langle \dot h^2 \rangle$ ($G_N$ is Newton's constant),  
 which can be expressed as  
$ \frac{\hbox{ $d E_{G}$}}{\hbox{ $d^3 x$}}= \frac{\pi}{2 G_N}
\int_{0}^{\infty}\! df f^2 S_h(f).$ \break
Therefore
$\rho_{G}(f)\!=\!\frac{\pi}{2G_N}f^3S_h(f)$ and finally
$
\sqrt{S_h(f)}=\sqrt{\frac{2 G_N}{\pi}}\frac{\sqrt{\rho_{G}(f)}}{f^{3/2}}.
$
For our class of models the dilaton-driven {\small GW} spectrum is 
given in eq.(\ref{ogwdd}) and  
therefore $\sqrt{S_h(f)}$ for that part of the spectrum is  constant  
\begin{equation}
\sqrt{S_h(f)}= \sqrt{\frac{2 G_N} {\pi}} \frac{\sqrt{\rho_{G}^{S}}} {f_S^{3/2}}.
\end{equation}
Now, all we need to do is to 
translate the coordinates of the end-point of the minimal spectrum from
the $(f,\rho_{G})$ plane to the $\left(f,\sqrt{S_h(f)}\right)$. We 
summarize the results in  figure 1 showing the  interesting
region in the $(f,\sqrt{S_h(f)})$ plane.  

The cryogenic resonant
detectors EXPLORER and NAUTILUS provide the best direct
 upper limit  on the existence of
a relic graviton background, $S_h^{1/2}\!<\! 6\times 10^{-22} 
{\rm Hz}^{-1/2}$, at $f=920$ Hz \cite{18}. This limit
is still too high to be significant for our background. 
An important  way of improving sensitivity for the detection of
our background is to perform
cross-correlation measurements between as many working {\small GW} detectors 
as possible.

\section*{Acknowledgments}
This research is supported in part by the Israel Science Foundation and by an Alon
grant. I would like to thank my collaborators Massimo Giovannini, Maurizio Gasperini
and Gabriele Veneziano and acknowledge the help of many experts, experimentalists 
and theorists, in the field of GW detection.

\end{document}